\documentclass[areview]{elsarticle}

\usepackage{hyperref}
\usepackage{bbding}
\usepackage[english]{babel}
\usepackage{multirow}
\usepackage{caption}
\usepackage{booktabs}
\usepackage{relsize}
\usepackage{array}
\usepackage{threeparttable}
\usepackage[textwidth=345pt]{geometry}
\usepackage{blindtext}
\usepackage{makecell}
\usepackage{amsthm,amsmath,amssymb,lipsum}
\usepackage{mathrsfs}
\usepackage{url}
\usepackage[utf8]{inputenc}
\DeclareUnicodeCharacter{03B8}{\ensuremath{\theta}}

\begin{document}
\begin{sloppypar}

\begin{frontmatter}

\title{An Overview of AI and Blockchain Integration for Privacy-Preserving\tnoteref{mytitlenote}}

\author{Zongwei Li, Dechao Kong, Yuanzheng Niu, Hongli Peng, Xiaoqi Li\textsuperscript{*}, Wenkai Li}
\address[mymainaddress]{School of Cyberspace Security, Hainan University, Haikou, China}

\nonumnote{Email: csxqli@gmail.com (Xiaoqi Li)}
\nonumnote{Zongwei Li, Dechao Kong, Yuanzheng Niu, Hongli Peng are co-first authors.}

\begin{abstract}
With the widespread attention and application of artificial intelligence (AI) and blockchain technologies, privacy protection techniques arising from their integration are of notable significance. In addition to protecting privacy of individuals, these techniques also guarantee security and dependability of data. This paper initially presents an overview of AI and blockchain, summarizing their combination along with derived privacy protection technologies. It then explores specific application scenarios in data encryption, de-identification, multi-tier distributed ledgers, and k-anonymity methods. Moreover, the paper evaluates five critical aspects of AI-blockchain-integration privacy protection systems, including authorization management, access control, data protection, network security, and scalability. Furthermore, it analyzes the deficiencies and their actual cause, offering corresponding suggestions. This research also classifies and summarizes privacy protection techniques based on AI-blockchain application scenarios and technical schemes. In conclusion, this paper outlines the future directions of privacy protection technologies emerging from AI and blockchain integration, including enhancing efficiency and security to achieve a more comprehensive privacy protection of privacy. 
\end{abstract}

\begin{keyword}
\texttt Artificial Intelligence; Blockchain Privacy Protection; Data Encryption; De-identification; Access Control
\end{keyword}

\end{frontmatter}



\section{Privacy Security in AI and Blockchain}

\subsection{Development of Blockchain Technology}
The Bitcoin blockchain system, which was developed by Nakamoto and introduced in November 2008 \cite{nakamoto2008bitcoin}, has garnered significant attention and extensive discourse internationally. The appreciation in Bitcoin's value has led to the widespread use of the term cryptocurrency in both industry and scholarly circles. As of 18 February 2023 \cite{Binance}, Bitcoin's circulating market capitalization was RMB 3.25 trillion, indicating its significant commercial value and the boundless potential of virtual currencies in the financial market. This observation also suggests that blockchain technology has the potential to generate substantial profits for the industry. Significantly, the increased monetary value of Bitcoin has instigated a fresh surge of enthusiasm for research and development. The initial stage of blockchain technology, commonly referred to as Blockchain 1.0, is characterized by the utilization of distributed ledgers. The emergence of Ethereum, a blockchain technology developed in 2014, marked a significant milestone in the era of Blockchain 2.0. The Ethereum platform has incorporated various pioneering technologies, including smart contracts \cite{li2021hybrid}. Blockchain 3.0 has led to the development of application platforms for the Internet of Things and smart healthcare, as evidenced by sources \cite{mukherjee2021blockchain}. Blockchain technology is presently undergoing significant transformations, marking its entry into the fourth generation of blockchain. This generation is focused on creating a reliable ecosystem and implementing blockchain technology in diverse sectors, including cultural and entertainment, communication infrastructure, and other related domains \cite{schaefer2019transparent}.\par

The categorization of blockchains is primarily based on their level of accessibility and control, with the three main types being public, private, and federated chains. The decentralized nature of public blockchains, such as Bitcoin and Ethereum, allows nodes to freely enter or exit the network, thereby promoting maximum decentralization. As exemplified by FISCO BCOS \cite{17}, Federated chains facilitate the implementation of smart contracts utilizing the Turing-complete language and offer homomorphic cryptography to ensure privacy protection. However, federated chains are partially decentralized Private blockchain networks, such as Antchain, regulate node permissions while offering faster transaction processing and lower transaction fees.\par

\begin{figure}[h]
    \centering
    \includegraphics[width=1.1\textwidth]{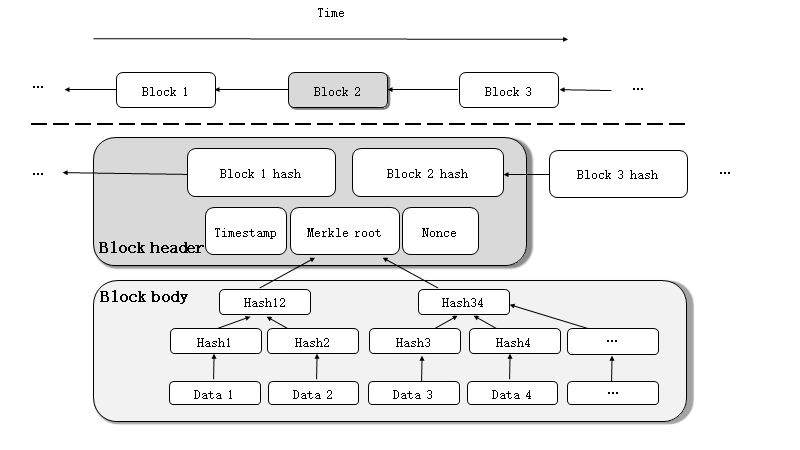}
    \caption{Ethereum Blockchain Structure}
    \label{figure1}
\end{figure}

The structure of the Ethereum blockchain is depicted in Figure \ref{figure1}. It utilizes a linked list data structure to establish connections between multiple blocks \cite{25}. The block header stores the hash address of the preceding block to establish a linkage between successive blocks. With the development and application of blockchain technology, the security concerns associated with blockchain technology in various fields can not be neglected. For instance, in the financial domain, the economic losses caused by privacy breaches are immeasurable \cite{li2022security}. The protection of user assets and identity information has emerged as a key area of focus in blockchain security research, given the criticality of these resources and the potential for malicious nodes to compromise them. Security is an issue that can not be separated from any industry or technology, and maintaining the security of the blockchain is essential to its future development.\par


Ethereum is a decentralized blockchain platform that employs multiple nodes to uphold a shared ledger of information collaboratively. Every node utilizes the Ethereum Virtual Machine (EVM) to compile smart contracts, and communication between nodes occurs via a peer-to-peer network \cite{de2020blockchain}. Nodes on Ethereum are provided with distinct functions and permissions. Nevertheless, all nodes can gather transactions and engage in block mining. When an Ethereum node acquires bookkeeping authority, it publishes a block, and the other nodes use the Proof of Stake (PoS) consensus mechanism to maintain data consistency.

It is worth noting that Ethereum exhibits a faster block generation speed than Bitcoin, with a block of roughly 15 seconds. It implies that miners can acquire block rewards quicker while the time interval for transaction verification is considerably diminished. Ethereum also supports smart contracts, allowing users to create digital wallets and DApps, among other applications.



\subsection{Artificial Intelligence}

Artificial intelligence (AI) \cite{zhang2023toward} is a field of study that focuses on developing computer systems that can simulate autonomous thinking and decision-making processes. AI exists to make the best judgments possible about actions and efficiently carry out predefined tasks. The capacity of DeepMind's Alpha Go to be the world champion and the success of OpenAI's ChatGPT model are just two examples of the phenomenal achievements that have helped AI attract substantial interest in recent years. Deep learning, natural language processing, and other areas can all be classified as subfields of artificial intelligence (AI). Although every subfield has a distinct collection of characteristics, they are all connected to the fundamental tasks of analyzing and parsing data.

Natural Language Processing (NLP) \cite{zhang2023toward} technology is a crucial area of AI that can deal with virtually any kind of text and linguistic data. Numerous top-notch NLP models, like the BERT model and GPT model, have emerged as a result of the development of deep learning models. NLP employs specific processing techniques or models to parse data with the aid of computers to do more study \cite{li2020stan}. Examples of downstream tasks in NLP include text categorization, speech recognition, and machine translation.


The first step of NLP is the text transformation that makes computers-understandable. However, this task is complicated by the complexity of language, ambiguity, and the need for evaluation criteria. To address these challenges, researchers often use specific symbolic forms and neural networks with different structures to perform various NLP tasks.

Deep Learning (DL) \cite{wasay2021deep} is increasingly prevalent in AI that is established by imitating the structure of neurons. By training hierarchical network structures, the external input information is processed layer by layer and then passes a hidden layer to form the final representation. DL is commonly categorized into supervised and unsupervised learning methods and has played a critical role in various fields such as image processing, speech recognition, and NLP.


\begin{figure}[ht]
    \centering
    \includegraphics[width=1\textwidth]{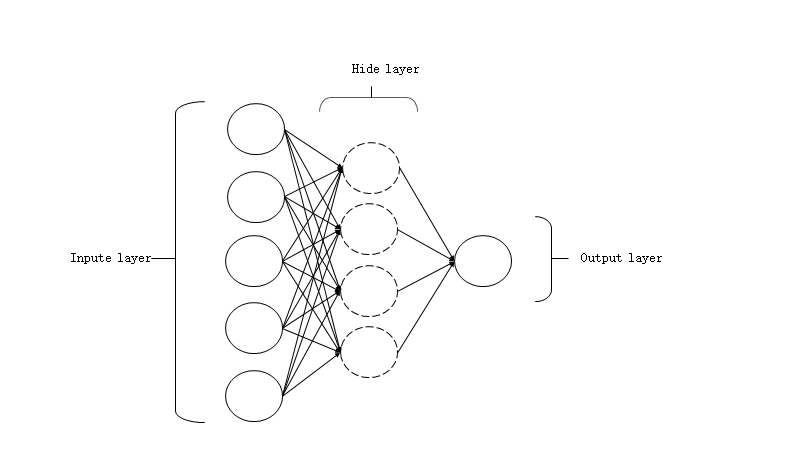}
    \caption{Deep Learning Perceptron Architecture}
    \label{figure2}
\end{figure}

As shown in Figure \ref{figure2}, DL employs a multi-level neural network structure to learn data features. The neural network contains three types of layers including the input layer, the hidden layer, and the output layer. Each perceptron layer is connected to the next to form the deep learning model. With the emergence of distributed deep learning, the computational parameters (e.g., eigenvalues, gradient values) during the model training have become crucial information transmitted among nodes. However, malicious nodes can potentially intercept computation results using advanced attack algorithms, leading to the exposure of sensitive data through reverse inference or even data leakage.

\subsection{Integration of Artificial Intelligence and Blockchain Technologies}
In the current era of information, artificial intelligence and blockchain technologies are increasingly being applied across various domains, with data security and privacy protection emerging as pressing issues to be addressed. Innovative cases, such as Anthropic's Constitutional AI \cite{Constitutional}, SingularityNET's Decentralized AI \cite{Distributed}, and ChainLink's Decentralized Oracle \cite{chainlink} are delving into the deep integration of artificial intelligence and blockchain technologies to achieve more efficient, secure, and transparent data processing.\par

Within Anthropic's Constitutional AI system, a combination of large-scale models and blockchain technologies ensures that audit tracking and accountability of model training data, parameters, and outputs are maintained. Similarly, SingularityNET's Decentralized AI system deploys AI models on blockchain networks, facilitating decentralized cooperation and services among models. This allows users to access models more conveniently, adjust model parameters, and acquire highly reliable services. Moreover, ChainLink's Decentralized Oracle system enables blockchain networks to securely access off-chain AI models and data sets while verifying their inputs and outputs, providing a trustworthy external information source for the blockchain.\par

These systems manifest the integration of artificial intelligence and blockchain technologies in the following aspects:\par

(1) Utilizing blockchain technology to store and record the parameters, training data, and inputs and outputs of models, ensuring transparency in model audits and accountability;\par
(2) Deploying AI models on blockchain networks to achieve decentralized cooperation and services among models, enhancing system stability and scalability;\par
(3) Facilitating secure access to external AI models and data through decentralized systems, enabling blockchain networks to acquire reliable external information;\par
(4) Leveraging blockchain-based incentive mechanisms and token designs to establish motivating connections and trust interactions between AI model developers and users.

\subsection{Privacy Protection in Blockchain}

Blockchain is an emerging distributed ledger technology that employs consensus algorithms, such as PoS \cite{LI202210378}, to maintain on-chain data, as well as encrypt transaction data through cryptography. Despite its transparency and openness, the exposure of sensitive data on the blockchain poses challenges to data privacy and security \cite{41}. In some use cases, such as financial distributed applications and supply chain projects, the disclosure of transaction data could lead to malicious competition and violation of established rules. To address these concerns and extend the applicability of blockchain, data security and privacy protection are critical issues that need to be addressed.

Various data protection technologies, such as zero-knowledge proof, ring signature, homomorphic encryption, and secure multi-party computing, have been proposed to enhance data privacy in the blockchain \cite{, li2020characterizing}. Among them, homomorphic encryption can be applied to data, as demonstrated in the FISCO BCOS of the federation chain. However, zero-knowledge proof and ring signatures are limited to user anonymity rather than data confidentiality.

\subsubsection{Zero-Knowledge Proof}

Zero-Knowledge Proof (ZKP) is a cryptographic technique proposed by Goldwasser et al. \cite{ni2023enabling} in 1985. It involves two roles, the prover and the verifier, with the prover demonstrating the correctness of a promise to the verifier without revealing any information about it. ZKP refers to a cryptographic method that allows a verifier to prove that a statement is true without knowing any other information. The ZKP protocol satisfies three properties: completeness, soundness, and zero-knowledge. The prover must provide sufficient evidence to convince the verifier of the promise's validity while revealing nothing except correctness. In the blockchain, zk-SNARKs is a commonly used algorithm to implement privacy due to its minimal verification computation and concise proofs \cite{2023802}. Zero-cash systems have employed zk-SNARKs to construct decentralized coin-mixing pools, where privacy is maintained through coin minting. The prover does not need to reveal which commitment belongs to them, thereby utilizing zk-SNARKs to encrypt personal privacy data.

\subsubsection{Ring Signature}

Ring signature (RS) \cite{liu2013linkable} was proposed by Rivest in 2001 in the context of anonymous checks. RS enables a signer to hide their own public key among a group of public keys, forming a ring of $n$ users who have equal status. Its application on the blockchain is to hide the transaction address of the trader. With a ring signature, even if the attacker collects the information of the ring members, he cannot infer the address of the trader. Users can freely choose members within the ring, and the more members in the ring, the greater the difficulty of the attack.

RS allows for untraceable signature schemes, where the signer uses the private and public keys of the ring members to sign and the verifier can only confirm that the signer is a member of the ring. And the algorithm process is as follows:
\begin{enumerate}
    \item[(1)]
    Generating public-private key pairs. By using the PPT algorithm and inputting the security parameters, users can obtain their public-private key pairs. Group members are capable of generating their public keys $PK$ and private keys $SK$ according to the PPT algorithm.
    \item[(2)]
    Signing. Based on the $message$ and the public keys of the group members $PK_{i}$, as well as the signer's private key $sk$, the RS can be conducted.
    \item[(3)]
    Verification. Verify whether the signature of the message is true according to the input, $B = verify(message,signature)$. If $B = 1$, it is true, and if $B = 0$, it is false.
\end{enumerate}

Encryption technology has enabled the development of various types of ring signatures to cater to different needs. These include threshold ring signatures, associated ring signatures, and other variations. Users can choose the most appropriate type of ring signature based on specific requirements.

\subsubsection{Homomorphic Encryption}

Homomorphic Encryption (HE) \cite{yang2020zero} is a fundamental component of cryptography that plays a vital role in the development of blockchain technology. Traditional encryption techniques provide confidentiality, security, and immutability but do not ensure data availability. In contrast, HE enables operations on encrypted data without revealing the plaintext. HE can be categorized into partial homomorphic encryption, which supports a limited number of data operations, and fully homomorphic encryption, which supports any number of operations. By encrypting data with a public key, users or third-party organizations can perform operations without disclosing the underlying data, which can only be decrypted using the corresponding private key. The process is as follows, where $m$ is the data that needs to be encrypted.
\begin{enumerate}
    \item[(1)]
    Use a public key to perform homomorphic encryption on the data, $M_{1} = E_{pk} (m_{1} ), M_{2} = E_{pk} (m_{2})$.
    \item[(2)]
    Use a specific function F(x) to operate on the homomorphic encrypted data without obtaining plaintext information, $F(M_{1}, M_{2} )$.
    \item[(3)]
    Use a private key to perform homomorphic decryption on the processed data, $D_{sk} (F(M_{1},M_{2} ))$.
\end{enumerate}

Different HE technologies may be required for different application scenarios in the blockchain \cite{zhang2022authros}. In the initial stage of project deployment, the security requirements for encryption are relatively low, and using high-security homomorphic encryption technology will waste more time and computing resources. As the project progresses, security requirements will increase, and low-security HE technology may no longer meet security guarantees. In addition, the entry and exit of nodes in public blockchains increase the complexity of the system and node updates.

\subsection{Artificial Intelligence Privacy Protection Technology}

With the advancement of AI technology, AI has permeated many spheres of life, however in areas containing sensitive personal data (such as prescription medications), a range of security measures are needed to safeguard confidential information. Novel language analysis and perception models, like the ChatGPT model, are constantly emerging. Furthermore, AI has applications across various domains, including but not limited to image identification, autonomous driving, and large-scale data processing. The process of training models is contingent upon the availability of data, which serves as the fundamental basis for contemporary deep learning. In certain domains, it is possible that the data under analysis may contain confidential personal information, such as medical prescription data. Ensuring the prevention of personal data leakage is among the challenges posed by AI. Privacy data encompasses various types of information, such as personal assets, medical records, and behavioral patterns. The potential harm to users resulting from the exploitation of their private data by malicious actors has garnered the interest of scholars in both academic and industrial settings. Machine learning commonly employs various privacy protection techniques, such as secure multi-party computation, and homomorphic encryption. The majority of privacy protection schemes for AI across various contexts rely on three fundamental methods.

\subsubsection{Secure Multi-party Computation}

Data privacy protection has received extensive attention in the discipline of AI, where secure multi-party computation (SMPC) plays a crucial part in the protection of privacy \cite{20231109}.SMPC is a cryptographic technique designed to allow multiple participants to securely process and compute their private data without exposing those data. Its core idea is that multiple participants perform encrypted communication, and the parties collaborate to securely calculate the data. Each participant only knows their own calculated value to protect data privacy and security. SMPC is a research hotspot in cryptography, and commonly used protocols include garbled circuits and secret sharing.

Garbled circuits \cite{20231109} can achieve secure calculation of almost any specific function by combining secure gate circuits. Via combining these secure gate circuits, the garbled circuit protocol \cite{yao1982protocols} is capable of secure computation of nearly any function. In contrast, the secret-sharing protocol offers a more straightforward strategy, based on the distribution of a secret into multiple parts among participants. Each participant is privy to only their input value, with all participants working jointly to calculate a function. Given a threshold, the secret may be pieced back together when the result is within the threshold range, thereby ensuring that only one party knows of its value.

\subsubsection{Differential Privacy}

Differential privacy, initially proposed by Dwork et al in 2006 \cite{dwork2006differential}, has been extensively researched in the academic community. This privacy protection technology is based on data distortion, aimed at preserving the original data while processing sensitive information. By setting an influencing factor for sensitive data, differential privacy can determine the degree of noise added to the data set, thereby determining the extent to which sensitive information can influence data queries.

The noise may increase the ambiguity of the data in many processes, such as training models, specific algorithms, and model output. Achieving the optimal balance between noise and data availability represents a significant challenge for differential privacy technology. Excessive noise increases the cost of data privacy, while inadequate noise fails to meet privacy requirements. Technological advancements have enabled the implementation of differential privacy technology with reduced data interference. For example, the ESA (Encode-Shuffle-Analyze) framework implemented the shuffling of user data and achieved complete anonymity of user data \cite{1640940056100-1760332955}.

\subsection{Summary}

This section delivers an overview of blockchain and AI technology and also covers their corresponding privacy protection techniques. Blockchain platforms such as Ethereum rely on multiple nodes to ensure data consistency, whereas AI technology simulates human cognition, facilitating autonomous behavior. This section also describes some of the privacy-preserving technologies on the blockchain, including zero-knowledge proof, ring signatures, and homomorphic encryption. Additionally, the application of privacy protection techniques such as secure multi-party computation and differential privacy is presented in the context of AI.


\section{Privacy Protection Through the Integration of AI and Blockchain Technologies}
At present, the data trust system suffers from limitations that compromise data transmission reliability \cite{19}. To address this challenge, blockchain technology can be utilized to establish a secure and dependable data storage and sharing system that fortifies data security and privacy protection \cite{li2021clue}. Specific applications of the integration of artificial intelligence and blockchain in privacy protection technology are presented in Table \ref{table 1}. By enhancing the integration and implementation of these technologies, the security and protective capacity of the current data trust system can be substantially improved.
\begin{table}[t]
\centering
\caption{Summary of the Applications of Blockchain in AI Privacy Protection}
  \footnotesize
\begin{tabular}{p{2.3cm} p{3.5cm} p{2.5cm} p{2.2cm}}
\toprule
\textbf{Author} & \textbf{Blockchain Technology} & \textbf{AI Technology} & \textbf{Security Mechanism} \\
\midrule
    Khan et al.\cite{16} & Consensus Protocol,\newline Digital Signature & Machine Learning & Decentralization\\
    
   Jennath et al.\cite{17} & Permissioned Blockchain, Cryptographic Signature & Machine Learning & De-identification\\
    
    Chang et al.\cite{18} & Anonymity,\newline Multi-Signature & Deep Learning & Privacy Protection Algorithm\\
    
    Wang et al.\cite{19} & Tamper-Resistance,\newline Smart Contract & Machine Learning & Smart contract\\
    
    Wang et al.\cite{20} & Consortium Blockchain,\newline Incentive Mechanism & Federated Learning & De-identification\\
    Lin et al.\cite{22} & Consortium Blockchain, Smart Contract,\newline Cryptocurrency & Edge AI & Smart Contract\\
    Durga et al.\cite{23} & Privacy Protection,\newline Encryption key & Federated Learning & Encryption Protection\\
    Alshehri et al.\cite{24}& Decentralization,\newline Heterogeneous Encryption, Digital Signature & AI & Encryption Protection\\
    Tang et al.\cite{25} & Trustworthiness,\newline Homomorphic Encryption, Differential Privacy & Machine Learning & Privacy Protection Algorithm\\
    \bottomrule
    \end{tabular}%
  \label{table 1}%
\end{table}%
\subsection{Data Encryption}
Traditional data storage and sharing methods are vulnerable to numerous security threats \cite{16}\cite{17}\cite{18} as they rely on centralized servers, which makes them easy targets for attackers. This vulnerability results in serious issues, such as data leaks and tampering. Traditional encryption methods are no longer sufficient to meet the increasingly urgent security needs \cite{23}\cite{24}.

To address these problems, privacy protection technology based on the combination of artificial intelligence and blockchain technology has emerged. Utilizing distributed encryption algorithms greatly enhances the security and privacy protection level of data.

Considering that many traffic and vehicle data may contain sensitive personal information, Wang et al. \cite{20} proposed a blockchain-based privacy-preserving federated learning (FL) scheme. The scheme improves the Multi-Krum technique and combines it with homomorphic encryption to achieve ciphertext-level model aggregation and model filtering, which can verify local models while achieving privacy protection. In this scheme, the Paillier homomorphic encryption technique \cite{21} is used to encrypt model updates, providing further privacy protection.\par
The Paillier algorithm is as follows:

(1) Generating the key. Choosing two random large prime numbers $p$ and $q$ such that $n_{0}$ and $\lambda $ satisfy Formula \ref{equation2} and Formula \ref{equation3}, respectively. Then select a value $g\in  B$ that satisfies Formula \ref{equation5} based on Formula \ref{equation4}.
\begin{small}
    \begin{equation}
    \setlength{\abovedisplayskip}{0pt}
    \setlength{\belowdisplayskip}{0pt}
    Paillier.Genkey()\to {(n_{0} ,g),(\lambda , \mu )}\label{equation1}
    \end{equation}
    \begin{equation}
    \setlength{\abovedisplayskip}{0pt}
    \setlength{\belowdisplayskip}{0pt}
    n_{0} = p*q\label{equation2}
    \end{equation}
    \begin{equation}
    \setlength{\abovedisplayskip}{0pt}
    \setlength{\belowdisplayskip}{0pt}
    \lambda =lcm(p-1,q-1)\label{equation3}
    \end{equation}
    \begin{equation}
    \setlength{\abovedisplayskip}{0pt}
    \setlength{\belowdisplayskip}{0pt}
    L(x)=(x-1)/n_{0}\label{equation4}
    \end{equation}
    \begin{equation}
    \setlength{\abovedisplayskip}{0pt}
    \setlength{\belowdisplayskip}{0pt}
    gcd(L(g^{\lambda } mod n_{0}^{2} ))\label{equation5}
    \end{equation}
\end{small}

(2)Encrypting. With Formula \ref{equation6}, the Paillier algorithm is applied.
\begin{small}
    \begin{equation}
    \setlength{\abovedisplayskip}{0pt}
    \setlength{\belowdisplayskip}{0pt}
    Paillier.Enc(m)\to c=g^{m}\cdot  r^{n_{0} } mod n_{0}^{2} \label{equation6}
    \end{equation}
\end{small}

(3)Decrypting. With Formula \ref{equation7}, the ciphertext $c$ less than $n_{0} $.
\begin{small}
    \begin{equation}
    \setlength{\abovedisplayskip}{0pt}
    \setlength{\belowdisplayskip}{0pt}
    Paillier.Dec(c)\to m=L(c^{\lambda } mod n_{0}^{2} ) /L(g^{\lambda } mod n_{0}^{2}) mod n_{0} \label{equation7}
    \end{equation}
\end{small}

In this context, $(n_{0}, g)$ represents the public key, while $(\lambda, \mu)$ denotes the private key. The plaintext $m < Z_{n}$, and the random number $r < n_{0}$ .

\subsection{De-identification}
\begin{figure}
    \centering
    \includegraphics[width=1\textwidth]{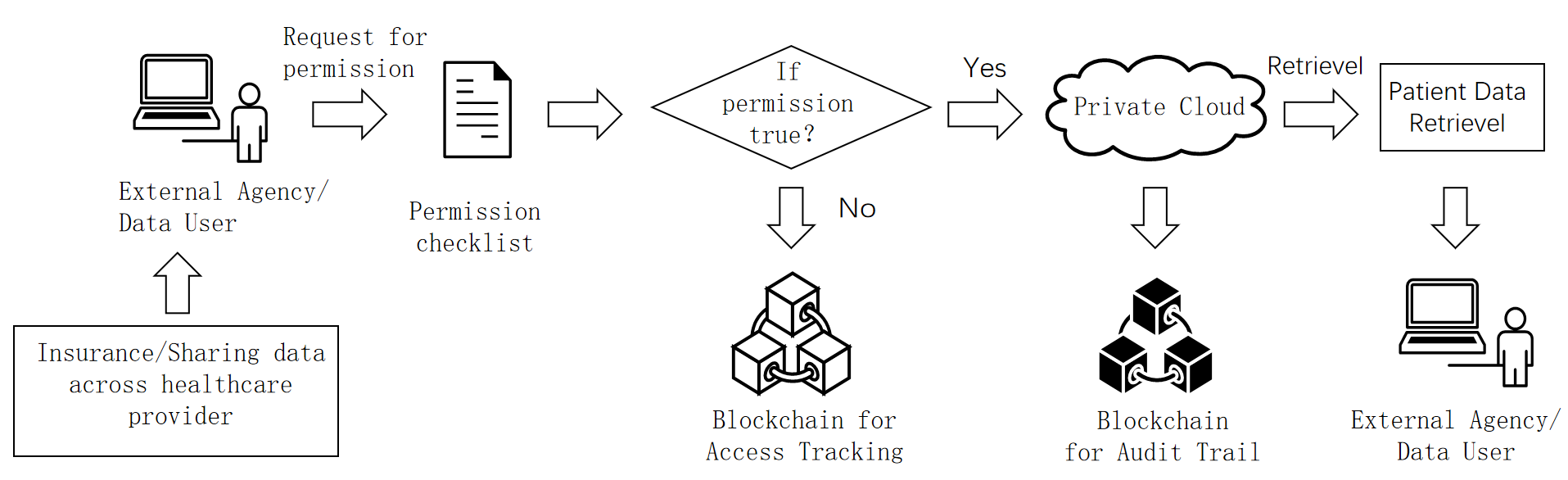}
    \caption{Patient Data Request Process}
    \label{figure 3}
\end{figure}

De-identification is a commonly used method for anonymizing personal identification information in data by separating data identifiers from the data itself, thereby reducing the risk of data tracking. Jennath et al.\cite{25} proposed a decentralized artificial intelligence framework based on permissioned blockchain technology that incorporates this approach. The framework effectively separates personal identification information from non-personal identification information and stores the hash value of personal identification information in the blockchain. This approach allows medical data to be shared without revealing patient identity. The framework employs two independent blockchains for data requests, as depicted in Figure \ref{figure 3}. One blockchain stores patient information and data access permission and the other captures audit traces of any queries or requests made by requesters. This design grants patients full ownership and control over their data while enabling secure data sharing among multiple entities.

\subsection{Multi-layered Distributed Ledger}
Multi-layer distributed ledger is a decentralized data storage system with multiple hierarchical layers designed to achieve efficient and secure data sharing and privacy protection\cite{18}\cite{25}.

Chang et al.\cite{18} proposed DeepLinQ, a blockchain-based multi-layer distributed ledger that addresses users' privacy concerns about sharing data by enabling privacy-protected data sharing. DeepLinQ leverages blockchain's characteristics, such as complete decentralization, consensus mechanism, and anonymity, to protect data privacy. It achieves this through several techniques, including on-demand querying, proxy reservation, subgroup signatures, access control, and smart contracts. By utilizing these techniques, DeepLinQ enables privacy-protected distributed data sharing.

\subsection{K-anonymity}

The K-anonymity method \cite{26} \cite{27} is a privacy protection technique that aims to group individuals in a dataset in such a way that each group contains at least K individuals with the same attribute values, thereby protecting the privacy of individuals. 

Long et al. \cite{26} proposed a reliable transactional model based on the K-anonymity method for transactions between electric vehicles and energy nodes. The K-anonymity method in this model serves two functions: first, to conceal user identifiers so that attackers cannot link users to their electric vehicles; second, to hide the location of electric vehicles by constructing a unified request using K-anonymity techniques to conceal the location of the car owner. 

\section{Evaluation and Situation Analysis}
This section presents a comprehensive evaluation and analysis of ten privacy protection systems proposed in recent years, which are based on the fusion of artificial intelligence and blockchain. The evaluation focuses on five key characteristics (authority management, access control, data protection, network security, and scalability), and discusses the strengths, weaknesses, and areas for improvement. The unique features of the fusion of artificial intelligence and blockchain, such as decentralization, distribution, and anonymity, provide new ideas and solutions for privacy protection \cite{49}. As shown in Table \ref{tab:sum}, multiple evaluation metrics are employed to derive the analytical results for the combined applications of AI and blockchain.\par

\begin{table}[t]
\centering
\caption{Evaluation Summary of Privacy Protection Systems Based on the Integration of Artificial Intelligence and Blockchain}
\label{tab:sum}
\begin{threeparttable} 
\relsize{-2} 
\begin{tabular}{>{\centering}m{2cm}>{\centering}m{4.8cm}ccccc}
\toprule
\textbf{Author}&\textbf{System Features}&\textbf{AM}&\textbf{AC}&\textbf{DP} &\textbf{NS}&\textbf{SC} \\
\midrule
        Singh et al.\cite{50}&A framework for IoT medical data privacy with FL and Blockchain.& \Checkmark  & \Checkmark & \Checkmark & \Checkmark & \Checkmark \\
        TURKI et al.\cite{51}&Smart parking system using AI and blockchain for privacy& \Checkmark  & \Checkmark & \Checkmark &\XSolid  & \XSolid \\
        Singh et al.\cite{52}&An AI-enabled blockchain-supported smart IoT framework& \XSolid  & \Checkmark & \Checkmark & \XSolid & \Checkmark \\
        Bosri et al.\cite{53}&AI blockchain-based privacy-preserving recommendation platform& \Checkmark  & \Checkmark & \Checkmark & \XSolid & \XSolid \\
        Elhoseny et al.\cite{54}&Blockchain and AI-based privacy-preserving framework& \XSolid  & \Checkmark & \Checkmark & \Checkmark & \Checkmark \\
        Wan et al.\cite{55}&Blockchain-based privacy framework for B5G edge devices& \Checkmark  & \Checkmark & \Checkmark & \XSolid & \Checkmark \\
        Deebak et al.\cite{56}&AI blockchain-based privacy framework for smart contracts& \Checkmark  & \Checkmark & \Checkmark & \XSolid & \Checkmark\\
        Wazid et al.\cite{57}&Blockchain-based secure communication for healthcare drones& \Checkmark  & \Checkmark & \Checkmark & \XSolid & \XSolid\\
        Leed et al.\cite{58}&Blockchain and AI-based privacy-preserving asset management& \Checkmark  & \Checkmark & \Checkmark & \XSolid & \Checkmark\\
        Ali et al.\cite{59}&Privacy and intrusion detection with blockchain and deep learning& \Checkmark&\Checkmark&\Checkmark&\Checkmark&\XSolid \\
\bottomrule
\end{tabular}
\begin{tablenotes}
\item In the presented table, the AM represents Authority Management, AC signifies Access Control, DP denotes Data Protection, NS represents Network Security, and SC indicates Scalability.
\end{tablenotes}
\end{threeparttable} 
\end{table}

\subsection{Authority Management}

Access control is a security technology that restricts user access to authorized resources based on predefined rules or policies, safeguarding system security and data integrity. Therefore, designing and implementing access control is an important issue in the research of blockchain and AI systems.\par

TURKI et al. \cite{51} developed an intelligent privacy parking management system based on AI and blockchain technology, which uses a Role-Based Access Control (RBAC) model to manage permissions \cite{52,53,54,60,61,62}. In this model, users are assigned one or more roles and classified according to roles to control attribute access permissions. Participants use their blockchain addresses to verify their identities and perform attribute authorization access.\par

Ali et al. \cite{59} developed a privacy protection and intrusion detection framework based on blockchain and deep learning, which defines four different levels of access control policies. The framework includes Membership Service Participants (MSPs) responsible for handling client registration and authentication processes, issuing and storing participant certificates. In addition, Key Verifiers (KVs) are responsible for verifying user identities and certificates requested by clients, using application programming interfaces to implement identity verification.\par

In addition, security measures such as multi-factor authentication \cite{52, 63} are also applied in some privacy protection systems based on the integration of artificial intelligence and blockchain. These security measures ensure that only authorized users can access relevant data in the system to achieve privacy data isolation. However, the access control design of many privacy protection systems based on the integration of artificial intelligence and blockchain is not well-developed. This study summarizes the possible reasons for access control defects and compiles them into Table \ref{tab:au}. The following summarizes the possible reasons for defects in these applications:\par

(1) Dependency on blockchain technology. Developers may focus more on technical implementation and performance, and may overlook details such as access control. Some developers may believe that blockchain technology itself has provided sufficient security, and therefore view access control as a secondary issue.\par

(2) Lacking of practical scenario testing. Access control requires an understanding of the system's usage scenarios and business logic. However, developers may focus on implementing business logic and lack consideration for practical scenarios.\par

Access control is an important component in protecting user privacy, and effective access control is required to avoid malicious behavior risks such as privacy breaches and data tampering.\par
\begin{table}[t]
\centering
\caption{Summary of Issues and Deficiencies in System Authority Management}
\label{tab:au}
\begin{threeparttable} 
\relsize{-2} 
\begin{tabular}{>{\centering}m{1.7cm}>{\centering}m{3.6cm}ccccccc}
\toprule
        {} & {} & \multicolumn{4}{c}{\textbf{Involved Features}}& \multicolumn{3}{c}{\textbf{Involved Layers}}\\
        \cmidrule(rl){3-6} \cmidrule(rl){7-9}
        \textbf{Types of Deficiencies} &\textbf{Causes of Deficiencies}& {IA} & {AM} & {AC} & {DP} & {CL}&{SL}&{AL} \\
        \midrule
        Authentication and authorization & Identity verification and authorization granting failed & \Checkmark & \Checkmark & \Checkmark & \XSolid&\Checkmark&\Checkmark&\Checkmark \\
        Permission leakage& Sensitive information leaked due to illegal access & \Checkmark & \XSolid & \Checkmark & \Checkmark&\Checkmark&\XSolid&\Checkmark \\
        Permission inheritance & Failure to inherit permissions from higher-level user & \Checkmark & \Checkmark & \Checkmark & \XSolid&\XSolid&\Checkmark&\Checkmark \\
        Unauthorized access & Illegally possessing higher-level permissions & \Checkmark & \Checkmark & \Checkmark & \Checkmark&\XSolid&\Checkmark&\Checkmark \\
       
        \bottomrule
        \end{tabular}
\begin{tablenotes}
\item In this table, IA represents Identity Authentication, AM denotes Authorization Management, AC is Access Control, DP expresses Data Protection, CL means Consensus Layer, SL points Smart Contract Layer, and AL signifies Application Layer.
\end{tablenotes}
\end{threeparttable} 
\end{table}

\subsection{Access Control}
Access control is one of the key mechanisms for privacy protection, restricting access based on user identity and group membership to ensure that only authorized users can access specific resources, thereby protecting the system from unauthorized access. To achieve effective access control, multiple factors need to be considered, such as user authentication \cite{51}, authorization \cite{58}, and access policies \cite{52}. Only when these aspects are adequately considered and implemented can privacy and security be ensured in the system.\par

Digital Identity Technology (DIT) \cite{50,51,52,58,63} is an emerging approach to IoT applications that can provide secure access control and protect device and data privacy. Wazid et al. \cite{57} proposed a series of access control policies based on digital identity technology and cryptographic primitives to protect communication security between entities such as drones, Ground Station Servers (GSS), and cloud servers. After the entity's registration is completed in the trusted control room, credentials are stored in memory and mobile devices. After user authentication and key establishment, the user sends a request to the GSS using a session key, and the GSS issues an encrypted command message with the current timestamp to the drone. The drone verifies its identity with the GSS and establishes a key between them for secure communication.\par

Lee et al. \cite{58} introduced blockchain-based access tokens to verify access control policies written in smart contracts and to use access tokens to authenticate Docker Registry, ensuring that only authorized users can access the models contained therein. These technologies can provide efficient and reliable secure access control mechanisms in IoT environments.\par

This study found that some systems still fail to implement effective access control. Table \ref{tab:ac} summarizes the types of defects and the involved layers. The potential reasons are as follows:\par

(1) Unreasonable system design. The system design process did not adequately consider privacy protection and access control or did not adequately consider whether strong identity authentication and access control are required based on the system's usage scenarios and requirements.\par

(2) Inadequate permission control. The granularity of access control is not fine enough, resulting in users obtaining unnecessary permissions and creating security vulnerabilities.\par

If access control mechanisms are lacking in the design and development of privacy protection systems based on the integration of artificial intelligence and blockchain, they will face security risks and threats of privacy breaches, endangering the stability and integrity of the system.\par

\begin{table}[t]
\centering
\caption{Summary of System Access Control Issues}
\label{tab:ac}
\begin{threeparttable} 
\relsize{-2} 
\begin{tabular}{>{\centering}m{1.7cm}>{\centering}m{3.7cm}ccccccc}
\toprule
        {} & {} & \multicolumn{4}{c}{\textbf{Involved Features}}& \multicolumn{3}{c}{\textbf{Involved Layers}}\\
        \cmidrule(rl){3-6} \cmidrule(rl){7-9}
        \textbf{Types of Deficiencies} &\textbf{Causes of Deficiencies}& {IA} & {AM} & {AC} & {DP} & {CL}&{SL}&{AL} \\
        \midrule
        Unauthorized access &Improper restriction of user access to resources &  \Checkmark &  \Checkmark &  \Checkmark &  \Checkmark& \Checkmark& \Checkmark& \Checkmark \\
        Revocation of privileges&Failure to promptly revoke expired privileges &  \Checkmark &  \Checkmark &  \Checkmark & \XSolid&\XSolid &\Checkmark& \Checkmark \\
        Unauthorized access & Users can access resources beyond their authorized scope &  \Checkmark &  \Checkmark &  \Checkmark &  \Checkmark&\XSolid& \Checkmark& \Checkmark \\
        Inappropriate strategy & Lack of appropriate access control policies in the system &  \Checkmark &  \Checkmark &  \Checkmark & \XSolid& \Checkmark& \Checkmark& \Checkmark \\
        Audit and monitoring&Failure to adequately audit and monitor access behavior& \XSolid & \XSolid &  \Checkmark &  \Checkmark& \Checkmark& \Checkmark& \Checkmark \\
       
        \bottomrule
        \end{tabular}
\begin{tablenotes}
\item In this table, IA represents Identity Authentication, AM denotes Authorization Management, AC is Access Control, DP expresses Data Protection, CL points Consensus Layer, SL means Smart Contract Layer, and AL signifies Application Layer.
\end{tablenotes}
\end{threeparttable} 
\end{table}

\subsection{Data Protection}
Data protection refers to measures such as access control, data encryption \cite{50,64}, data backup, and security auditing to ensure that user data is not illegally accessed, tampered with, or leaked. In terms of data processing, technologies such as anonymization \cite{52}, data masking \cite{56}, data encryption, and data isolation \cite{65} can be used to protect data from leakage and unauthorized access. In addition, encryption technologies such as differential privacy protection \cite{55}, homomorphic encryption \cite{59}, hash algorithms \cite{52,57}, digital signature algorithms \cite{59}, asymmetric encryption algorithms \cite{56,58} can ensure data confidentiality and prevent unauthorized access by non-authorized users.\par

Wan et al. \cite{55} used homomorphic encryption to encrypt the local model parameters of edge devices and uploaded them to the central server. The central server aggregates ciphertext parameters from all clients and sends the updated global model back to each client for decryption to complete the training process. In addition, this study used differential privacy protection to protect the privacy of local model parameters. Differential privacy adds noise data, such as Laplace noise \cite{66}, to the response to prevent privacy data leakage.\par

Lee et al. \cite{58} proposed a management framework that uses blockchain and AI technology to protect the privacy of digital assets. This study uses OrbitDB as an off-chain distributed database to store personal data of end-users. Before providing services, permission to the personal OrbitDB database of the end-user is required. The end-user uses asymmetric encryption algorithms to encrypt the OrbitDB address with public key $PK_{enc}$ and encrypts the data stored in OrbitDB with public key $PK_{data}$ to ensure that the data is not readable in plaintext, ensuring data security and privacy, and preventing data abuse.\par

Despite the widespread use of encryption technology for data security, it only serves as a partial safeguard for data protection. Therefore, it is imperative to adopt multiple security technologies to establish a comprehensive security system to ensure data security. As depicted in Table \ref{tab:dp}, we summarize the issues related to data protection in the system. Some systems lack effective local or third-party security mechanisms, which may be due to the following reasons: \par
(1) Security relies heavily on the blockchain, thereby neglecting other cryptographic measures during the design process.\par 
(2) Difficulties in designing encryption algorithms based on real-world scenarios, which require designers to have a deep understanding of cryptographic algorithms and practical scenario demands. \par

Therefore, careful consideration is required when designing and using systems, and continuous attention and research on the development of cryptographic algorithms and cryptographic technology is necessary. The adoption of a combination of multiple technologies is essential to ensure the confidentiality, integrity, and reliability of data and to protect user privacy.\par

\begin{table}[t]
\centering
\caption{Summary of Data Protection Issues}
\label{tab:dp}
\begin{threeparttable} 
\relsize{-2} 
\begin{tabular}{>{\centering}m{1.6cm}>{\centering}m{3.7cm}ccccccc}
\toprule
        {} & {} & \multicolumn{4}{c}{\textbf{Involved Features}}& \multicolumn{3}{c}{\textbf{Involved Layers}}\\
        \cmidrule(rl){3-6} \cmidrule(rl){7-9}
        \textbf{Types of Deficiencies} &\textbf{Causes of Deficiencies}& {DP} & {AC} & {ET} & {CM} & {CL}&{SL}&{AL} \\
        \midrule
        Data leakage &Vulnerabilities or unauthorized access exist in the system & \Checkmark & \Checkmark & \Checkmark &\XSolid&\Checkmark&\XSolid&\Checkmark \\
        Data integrity &Data in the system has been maliciously tampered with & \Checkmark &\XSolid & \Checkmark & \Checkmark&\Checkmark&\Checkmark&\Checkmark \\
        Privacy leakage &Malicious users obtained sensitive information & \XSolid & \Checkmark & \Checkmark & \XSolid&\Checkmark&\XSolid&\Checkmark \\
        Data backup &Insufficient or untimely data backup in the system & \Checkmark & \XSolid & \XSolid & \XSolid&\Checkmark&\XSolid&\Checkmark \\

        \bottomrule
        \end{tabular}
\begin{tablenotes}
\item In this table, DP represents Data Protection, AC denotes Access Control, ET is Encryption Technology, CM expresses Consensus Mechanism, CL is Consensus Layer, SL points Smart Contract Layer, and AL means Application Layer.
\end{tablenotes}
\end{threeparttable} 
\end{table}

\subsection{Network Security}
The field of network security encompasses multiple aspects such as preventing network attacks, ensuring data confidentiality and integrity, and protecting systems from malicious software and network viruses. To ensure system security and reliability, a series of security measures and secure network architectures and protocols need to be adopted \cite{68}. Additionally, assessing and analyzing various network threats and developing corresponding security strategies and defense mechanisms are necessary to improve the security and reliability of the system.\par

Ali et al. \cite{59} proposed a privacy-preserving and intrusion detection framework based on blockchain and deep learning. This framework includes an Intrusion Detection System (IDS) that monitors and analyzes network traffic. IDS detects intrusion using feature-based and anomaly-based methods. The former matches rules/signatures with a database of known rules, while the latter relies on interpreting normal activity to identify any deviations. Deep learning algorithms automatically reduce the complexity of network traffic to identify correlations between data, effectively detecting intrusion behavior.\par

Singh et al. \cite{50} proposed an IoT healthcare privacy protection framework that combines federated learning and blockchain technology. This framework enhances security by introducing protocols in IoT devices and intelligent systems. These protocols can exchange initial control messages to ensure features such as message authentication, complete forward secrecy, and prevention of replay attacks, thereby preventing eavesdroppers from attempting to invade patient privacy. Table \ref{tab:nw} summarizes the types and features of network attacks that may be carried out against privacy protection systems based on the integration of artificial intelligence and blockchain. For systems that have not adopted effective network security measures, Table \ref{tab:nw} provides relevant design characteristics and involves system-level measures to prevent the following three scenarios: over-reliance on the security of the blockchain network, ignoring the design and implementation of security protocols, and prioritizing functionality and performance at the expense of security.\par

The absence of network security protection in a system may result in various security threats and risks, potentially leading to sensitive data and privacy information leakage and misuse, and causing significant losses to individuals and businesses. Therefore, network security protection is crucial in privacy protection systems based on the integration of artificial intelligence and blockchain and must be given adequate attention and protection.\par

\begin{table}[t] 
\centering
\caption{Summary of Network Attacks on the Systems}
\label{tab:nw}
\begin{threeparttable} 
\relsize{-2} 
\begin{tabular}{>{\centering}m{4.5cm}cccccccc}
\toprule
        {}  & \multicolumn{5}{c}{\textbf{Involved Features}}& \multicolumn{3}{c}{\textbf{Involved Layers}}\\
        \cmidrule(rl){2-6} \cmidrule(rl){7-9}
        \textbf{Types of Attacks} & {DP} & {IA} & {ENC} & {LB}&{DI} & {NL}&{CL}&{AL} \\
        \midrule
        Network sniffing attack &\Checkmark & \Checkmark & \Checkmark &\XSolid & \XSolid&\Checkmark&\XSolid&\Checkmark \\
        Distributed denial-of-service (DDoS) attack &\XSolid & \XSolid & \XSolid & \Checkmark & \XSolid&\Checkmark&\Checkmark&\Checkmark \\
        Man-in-the-middle (MITM) attack &\Checkmark & \Checkmark & \Checkmark & \XSolid & \XSolid&\Checkmark&\Checkmark&\Checkmark \\
        Malware attack &\Checkmark& \Checkmark & \Checkmark & \XSolid & \XSolid&\Checkmark&\Checkmark&\Checkmark \\
        Data tampering attack&\Checkmark& \Checkmark & \Checkmark & \XSolid & \Checkmark&\Checkmark&\Checkmark&\Checkmark \\
       
        \bottomrule
        \end{tabular}
\begin{tablenotes}
\item In this table, DP represents Data Protection, IA denotes Identity Authentication, ENC indicates Encryption, LB means Load Balancing, DI is Data Integrity, NL points Network Layer, CL expresses Consensus Layer, and AL signifies Application Layer.
\end{tablenotes}
\end{threeparttable} 
\end{table}

\subsection{Scalability}
The scalability refers to the ability of a system to handle an increasing number of users or larger amounts of data. When designing a scalable system, considerations must be given to system performance, node management, data storage, and transmission, among other factors. While ensuring system scalability, system security must also be taken into account to prevent security risks and data breaches.\par

Lee et al. \cite{58} designed a system that complies with the European General Data Protection Rules (GDPR) by storing artwork metadata and privacy-related data in a distributed file system off the chain. Digital tokens and artwork metadata are stored in OrbitDB, a database that stores data on multiple nodes to ensure data security. The off-chain distributed system improves system scalability by dispersing data storage.\par

Wan et al. \cite{55} proposed a blockchain-based B5G edge device privacy protection framework that uses federated learning to achieve distributed learning of local data. The central server aggregates encrypted local parameters from all clients and updates the global model. The use of blockchain technology decentralizes the federated learning server, reducing the risks of single-point failure and poisoning attacks. Additionally, the framework can be applied to different datasets, models, computing resources, and algorithms, improving system scalability. Furthermore, the framework can improve model interpretability and better handle bias and noise.\par

This study found that most systems lack sufficient design in terms of scalability or overly rely on the distributed nature of blockchain. Blockchain has several scalability issues, including scaling problems, low transaction processing speeds, and interoperability issues. Table \ref{tab:sc} lists the challenges and difficulties faced when designing systems with good scalability.\par

To improve system scalability, techniques such as distributed storage, distributed computing, data sharding, and parallel processing can be employed. In privacy protection systems based on the integration of artificial intelligence and blockchain, due to the processing of large amounts of sensitive data, the system needs to continuously expand to meet user demand. Therefore, when designing a privacy protection system, scalability must be considered to ensure continuous and stable operation. InterPlanetary File System (IPFS) \cite{58,67}, federated learning \cite{50,52,69,70}, data sharding technology \cite{71}, and parallel computing \cite{72} can all be used to improve system scalability.\par

\begin{table}[t]
\centering
\caption{Summary of System Scalability Issues}
\label{tab:sc}
\begin{threeparttable} 
\relsize{-2} 
\begin{tabular}{>{\centering}m{1.5cm}>{\centering}m{3.7cm}ccccccc}
\toprule
        {} & {} & \multicolumn{4}{c}{\textbf{Involved Features}}& \multicolumn{3}{c}{\textbf{Involved Layers}}\\
        \cmidrule(rl){3-6} \cmidrule(rl){7-9}
        \textbf{Types of Deficiencies} &\textbf{Causes of Deficiencies}& {SC} & {PF} & {AV} & {DC} & {NL}&{CL}&{SL} \\
        \midrule
        Throughput &Transaction processing throughput insufficient for high-concurrency requirements & \Checkmark & \Checkmark & \Checkmark & \XSolid&\Checkmark&\Checkmark&\Checkmark \\
        Scalability &The system size cannot meet the increased demand & \Checkmark & \Checkmark & \Checkmark & \XSolid&\Checkmark&\Checkmark&\Checkmark \\
        Load balancing &Imbalanced load on nodes in the system & \Checkmark & \Checkmark & \Checkmark & \XSolid&\Checkmark&\Checkmark&\Checkmark \\
        Data consistency &Data inconsistency exists across different nodes & \Checkmark & \XSolid & \Checkmark & \Checkmark&\Checkmark&\Checkmark&\Checkmark \\
       
        \bottomrule
        \end{tabular}
\begin{tablenotes}
\item In this table, SC represents Scalability, PF denotes Performance, AV indicates Availability, DC is Data consistency, NL means Network Layer, CL expresses Consensus Layer, and SL points Storage Layer.
\end{tablenotes}
\end{threeparttable} 
\end{table}

\subsection{Situation Analysis}
The integration of blockchain technology and AI has resulted in a system that effectively protects user privacy data. While this system still faces challenges such as data protection, access control, network security, and scalability, it is crucial to comprehensively consider and weigh these issues during the design phase based on practical considerations. As technology advances and application scenarios expand, this privacy protection system based on blockchain and AI is poised to receive widespread attention and research in the future.\par Based on the aforementioned research findings, application scenarios, and technical approaches, we can classify them into three categories:\par

(1) Privacy protection applications in the Internet of Things (IoT) utilizing both artificial intelligence and blockchain technology;\par
(2) Privacy protection applications in smart contracts and services that merge artificial intelligence and blockchain technology;\par
(3) Large-scale data analysis techniques that integrate artificial intelligence and blockchain technology for privacy protection.\par

Firstly, the technologies in the first category focus on how to apply blockchain and artificial intelligence for privacy protection in IoT devices. These studies employ artificial intelligence to analyze voluminous data, while taking advantage of the immutable and decentralized features of the blockchain to ensure data security and authenticity. This direction can handle massive IoT data, improve data processing efficiency using artificial intelligence, and guarantee data security with blockchain technology. However, the complexity of the processing framework leads to higher system development and maintenance costs, and the need to manage ever-growing data volumes.\par

Secondly, the technologies in the second category aim to fuse blockchain and artificial intelligence technology for privacy protection within smart contracts and services. These studies combine data processing and analysis with artificial intelligence, along with blockchain technology to record transactions and reduce dependency on trusted third parties. This direction secures interactions between users and services without revealing sensitive information and enhances the system performance using artificial intelligence. Nevertheless, this framework may encounter compatibility issues on existing infrastructure and result in elevated development and maintenance costs.\par

Lastly, the technologies in the third category concentrate on utilizing blockchain and artificial intelligence techniques in large-scale data analytics to achieve privacy protection. These studies exploit the blockchain's immutability and decentralization properties to ensure data security and authenticity, while harnessing artificial intelligence technology to enhance the accuracy of data analysis. The advantages of this direction include the ability to securely conduct large-scale data analysis while preserving privacy, the reduction of data authenticity issues present in traditional data analysis methods using blockchain, and the deep exploration of data with artificial intelligence technology to boost the quality of the analysis. However, the implementation of privacy protection techniques may increase the complexity of data analysis, and striking a balance between public and private aspects of the blockchain remains a topic of investigation.\par

\section{Conclusion and Prospect}

This work primarily investigates the application scenarios of privacy protection technologies fused with artificial intelligence and blockchain, introducing their related methodologies while evaluating five crucial characteristics. Moreover, it delves into the deficiencies within current systems and offers recommendations for improvement. Finally, these technologies are classified and summarized based on application scenarios and technical solutions. This study holds referential value for the development of AI and blockchain fusion and provides novel insights and directions for future research.\par

Challenges remain to be addressed in the realm of privacy protection technologies built upon AI and blockchain fusion, such as striking a balance between privacy preservation and data sharing. Exploring the fusion of AI and blockchain privacy protection technologies remains a worthwhile research direction. Consequently, we summarize several ways to integrate other techniques:\par

(1) Edge computing. By leveraging edge devices to process private data, edge computing achieves decentralization. Due to AI processing necessitating substantial computational resources, incorporating edge computing enables the distribution of computational tasks to edge devices for processing. This reduces transmission latency and network congestion while enhancing system processing speed and performance.\par

(2) Multi-chain mechanisms. Multi-chain mechanisms have the capacity to resolve single-chain blockchain performance and storage issues, thereby bolstering system scalability. The integration of multi-chain mechanisms allows for data classification based on distinct attributes and privacy levels, consequently improving the security and storage capabilities of privacy protection systems.\par

\footnotesize
\bibliography{mybibfile}

\end{sloppypar}
\end{document}